\documentclass[journal]{IEEEtran}
\usepackage{amsmath,amsfonts}
\usepackage{algorithmic}
\usepackage{algorithm}
\usepackage{array}
\usepackage[caption=false,font=normalsize,labelfont=sf,textfont=sf]{subfig}
\usepackage{textcomp}
\usepackage{stfloats}
\usepackage{url}
\usepackage{verbatim}
\usepackage{graphicx}
\usepackage{cite}
\begin{document}

\title{Ring Artifacts Correction Based on Global-Local Features Interaction Guidance in the Projection Domain}

\author{Yunze Liu\textsuperscript{†}, Congyi Su\textsuperscript{†}, Xing Zhao\textsuperscript{*}%
    \thanks{This work was supported the National Key Research and Development Program of China (No. 2020YFA0712200, No. 2023YFA1011402) and the National Natural Science Foundation of China (No. 12426308). }
    \thanks{Yunze Liu and Congyi Su are with the school of Mathematical Sciences, Capital Normal University, 105 West Third Ring Road North, 100048, Beijing, China. (email: 2220502075@cnu.edu.cn, 2230502043@cnu.edu.cn).}
    \thanks{Xing Zhao is with the school of Mathematical Sciences, Capital Normal University and National Key Laboratory of Strength and Structural Integrity, Aircraft Strength Research Institute of China. (email: zhaoxing\_1999@126.com).}
    \thanks{\textsuperscript{*}Corresponding author.}
    \thanks{\textsuperscript{†}Yunze Liu and Congyi Su contributed equally to this work.}}

%

\maketitle
\begin{abstract}
Ring artifacts are common artifacts in CT imaging, typically caused by inconsistent responses of detector units to X-rays, resulting in stripe artifacts in the projection data. Under circular scanning mode, such artifacts manifest as concentric rings radiating from the center of rotation, severely degrading image quality. In the Radon transform domain, even if the object's density function is piecewise discontinuous in certain regions, the projection images remain nearly continuous in the angular direction, making the ideal projections exhibit a smooth global low-frequency characteristic. In practical scanning, the local disturbances of the same detector unit at different scanning angles lead to a prominent high-frequency locality of stripe artifacts. Existing studies generally model ring artifacts disturbances as fixed additive errors, which overlooks the dynamic variation of detector responses during practical scanning. However, the degree of detector response inconsistency is a function of the projection values, as revealed in our experiments, thereby requiring consideration of the interaction between global and local features in the process of stripe artifacts extraction and correction. Therefore, we propose a CT ring artifacts correction method based on global and local features in the projection domain. We employ the VSS block and Dense block to respectively correct the low-frequency sub-band, which capture the global correlations of the projection, and the high-frequency sub-band, which contain local stripe artifacts after wavelet decomposition. Specifically, the accuracy of artifacts correction is enhanced by the interaction guidance between global and local features. Extensive experiments demonstrate that our method achieves superior performance in both quantitative metrics and visual quality, verifying its robustness and practical applicability.

\end{abstract}


\section{Introduction}
\begin{figure}[t]
  \centering
  \includegraphics[width=3.4in]{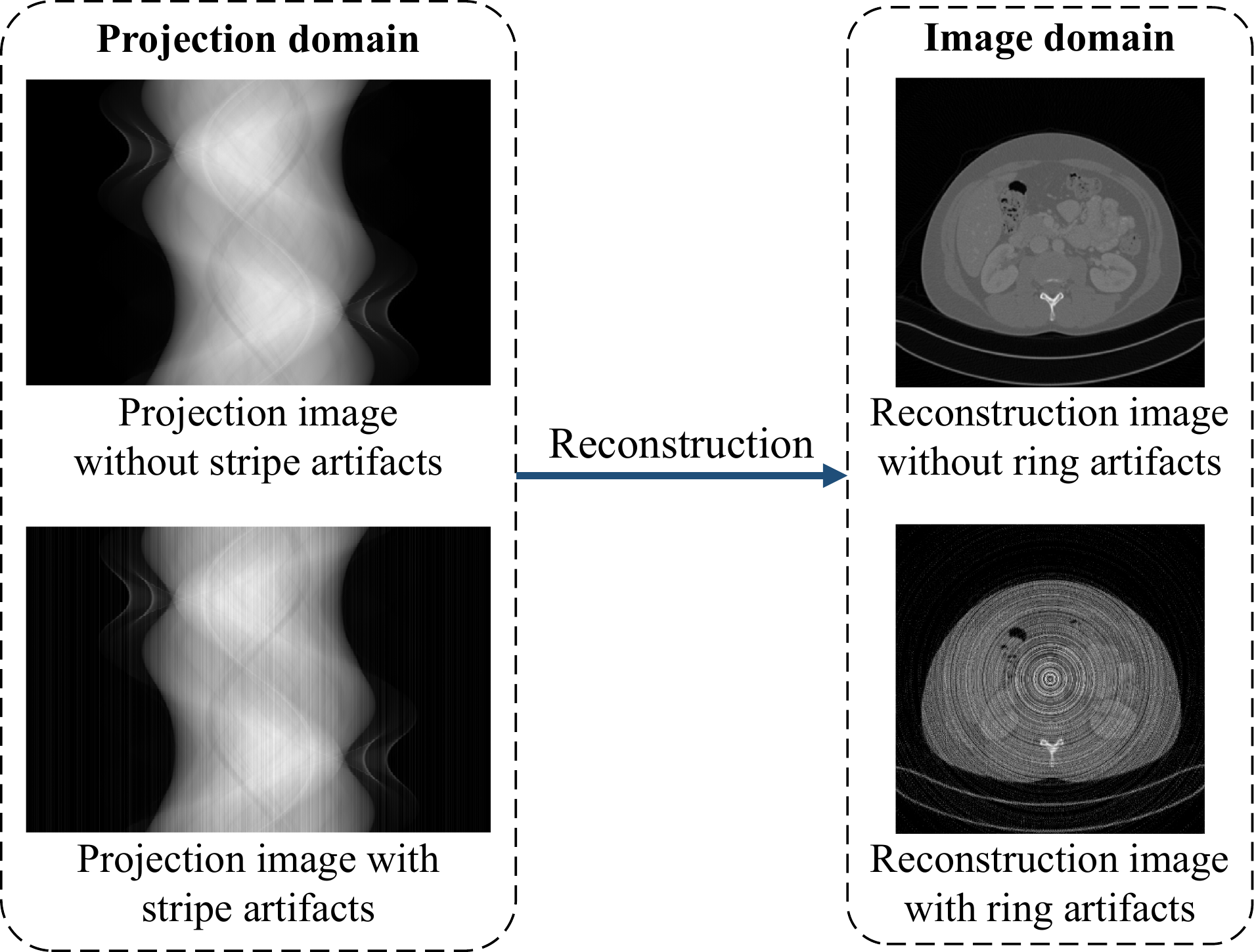}
  \caption{Characterization of ring artifacts in CT image and projection domains.}\label{fig1}
\end{figure}

Ring artifacts in CT images mainly arise from the inconsistent responses of detector units to X-ray. These inconsistencies caused by factors such as variations in scintillation screen thickness, non-uniform detector sensitivity, or amplifier circuit fluctuations~\cite{bornefalk2013effect}. These inconsistencies lead to significant deviations in recorded values even under identical X-ray exposure. When projection data are arranged in angular order, such errors appear as stripe artifacts along the detector direction, as illustrated in the projection domain of Fig.~\ref{fig1}. After reconstruction, these stripe artifacts appear as multiple concentric rings centered at the image center, with increasing intensity toward the center, as shown in the image domain of Fig.~\ref{fig1}. These artifacts obscure structural details and compromise the accuracy of lesion diagnosis, defect detection, and material analysis in CT imaging. Therefore, effective correction of ring artifacts has become a critical topic for improving CT image quality.\par
\IEEEpubidadjcol
Currently, ring artifacts correction methods in CT imaging can be broadly categorized into image-domain, projection-domain, and dual-domain methods. Image-domain methods typically transform ring artifacts into stripe artifacts via polar coordinate conversion for easier correction, followed by inverse transformation to the Cartesian domain. Representative methods include segmentation-based~\cite{shuo2014new,sijbers2004reduction,brun2009improved}, transform-based~\cite{zhang2012ring,wei2013ring}, decomposition-based~\cite{chen2009ring,yang2020post,li2021sparse}, and total variation (TV)-based techniques~\cite{2015LinRing,huo2016removing,liang2017iterative,wu2019removing,yan2016variation}. However, interpolation errors induced by coordinate conversion often degrade image resolution and introduce secondary artifacts. In recent years, deep learning-based image-domain correction methods have emerged. For instance, Hein \textit{et al.} employed a U-Net architecture with spectral-VGG loss~\cite{hein2023ring}, while Zhao \textit{et al.} developed Smooth GAN using adversarial learning with smoothness constraints~\cite{zhao2018removing}. Nevertheless, these methods are limited by insufficient generalization, structural distortion, and detail loss.

Projection-domain methods often utilize frequency decomposition for artifacts suppression. A typical example is the Wavelet-Fourier Filtering (WFF) method~\cite{munch2009stripe}, where Discrete Wavelet Transform (DWT) is used for high-frequency and low-frequency separation, followed by frequency-domain filtering. Guo \textit{et al.} further improved this method by introducing weighted averaging and damping filters~\cite{guo2015ct}. Deep learning has also been introduced for projection-domain correction. Nauwynck \textit{et al.} proposed a U-Net-based network with a weighted loss leveraging projection sparsity~\cite{nauwynck2020ring}. Fu \textit{et al.} combined residual learning and transfer learning, using wavelet subbands to separate artifacts while preserving details~\cite{fu2023deep}.\par

Dual-domain methods combine both image and projection priors within an optimization framework~\cite{paleo2015ring,salehjahromi2019new,zhu2024dual}. Moreover, Fang \textit{et al.} proposed a five-stage U-Net to perform domain-specific correction, integrating image, projection, and polar domain processing with parallel training for improved artifact suppression~\cite{fang2020removing}.

The Radon transform, proposed by Johann Radon in 1917, provides the mathematical foundation for projection acquisition and image reconstruction in CT imaging~\cite{radon1986determination,kak2001principles}. It describes the process of acquiring projection data by integrating a two-dimensional image along a set of specified lines. Moreover, Wu \textit{et al.}~\cite{wu2013continuity} demonstrated that the Radon transform remains continuous for almost all scanning angles, even when the object's density function exhibits boundary or jump discontinuities. \textbf{This theoretical result indicates that the ideal projection data should exhibit well smoothness and global consistency.}\par
\begin{figure*}[t]
  \centering
  \includegraphics[width=\textwidth]{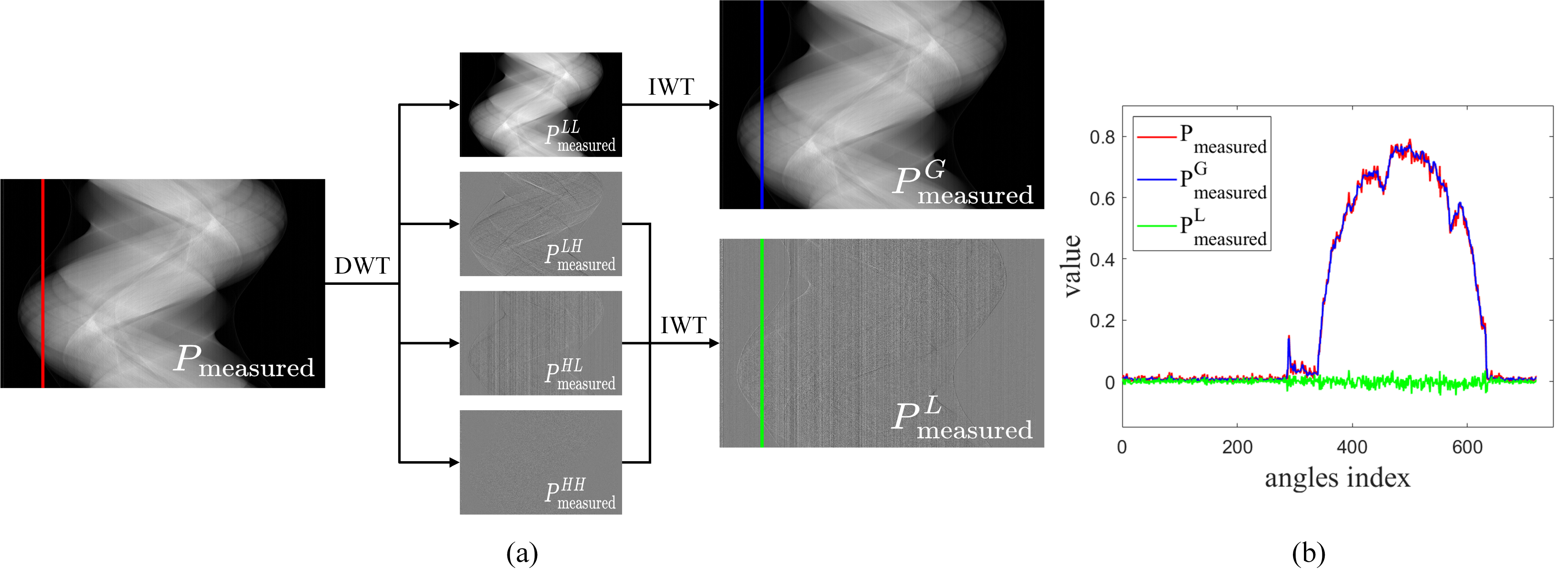}
  \caption{(a) Illustration of the inverse discrete wavelet reconstruction using the vertical high-frequency sub-band and the low-frequency sub-band from a real projection image. (b) Profile curve along the red, blue, and green lines shown in (a).}\label{interact}
\end{figure*}

In fan-beam CT scanning, detector response inconsistency arises from both intrinsic hardware variation and differences in ray paths through the object. \textbf{This leads to angle-dependent variations in the response of a single detector unit, resulting in localized stripe artifacts in the projection data.} To model this effect, several studies~\cite{salehjahromi2019new,liang2017iterative,wu2019removing} approximate the corrupted projection as the sum of the ideal projection and a fixed additive error term, which can be formulated as:
\begin{equation}
P'_i = P_i + G_i
\end{equation}
where $P_i$ denotes the ideal projection, $P'_i$ represents the corrupted projection, and $G_i$ is the additive error term associated with the $i$-th detector unit.

In this case, the projection image with stripe artifacts is modeled as the sum of a constant additive term $G_i$ and the ideal projection $P_i$. However, this formulation neglects the angle-dependent variability of detector response, limiting the effectiveness of additive-model-based correction methods.\par
To investigate this variability of a single detector unit, we applied DWT to a measured projection image $P_{\text{measured}}$, obtaining a low-frequency sub-band image $P^{LL}_{\text{measured}}$ and three high-frequency sub-band images in horizontal, vertical, and diagonal directions: $P^{LH}_{\text{measured}}$, $P^{HL}_{\text{measured}}$, and $P^{HH}_{\text{measured}}$, respectively. Owing to the geometric characteristics of stripe artifacts, DWT facilitates their localization and separation. As shown in Fig.~\ref{interact} (a), the vertical high-frequency sub-band effectively captures the localized features of stripe artifacts, while the low-frequency sub-band retains the global structure of the projection image. We further reconstructed the low-frequency and high-frequency sub-band images separately to obtain a global image $P_{\text{measured}}^{G}$ and a local image $P_{\text{measured}}^{L}$. Then, we extracted the values of the same detector unit across different projection angles from $P_{\text{measured}}$, $P^G_{\text{measured}}$, and $P^L_{\text{measured}}$. These responses are annotated with red, green, and blue lines, respectively in Fig.~\ref{interact} (a), and their corresponding value curves are plotted in Fig.~\ref{interact} (b). The results indicate that the inconsistency in detector response varies across projection angles: it becomes significantly more pronounced when rays pass through the object, while remaining relatively minor in regions not intersecting with the object.\par

Therefore, inconsistent responses of the detectors should be modeled as a projection-dependent functional perturbation rather than a fixed additive term, formulated as:
\begin{equation}
    P' = G_i(P)
\end{equation}
where $f_i$ denotes the response inconsistency function of the $i$-th detector unit, $P$ is the ideal projection image, and $P'$ is the projection image with stripe artifacts. For different detector unit $i$ and $j$, their response characteristics are assumed to be independent, i.e., $f_i\neq f_j$. \textbf{Under this modeling assumption, effective stripe artifact correction requires not only separate modeling of local and global features but also the interaction between them, enabling more accurate artifact extraction and correction.}\par
\begin{figure}[ht]
  \centering
  \includegraphics[width=3.3in]{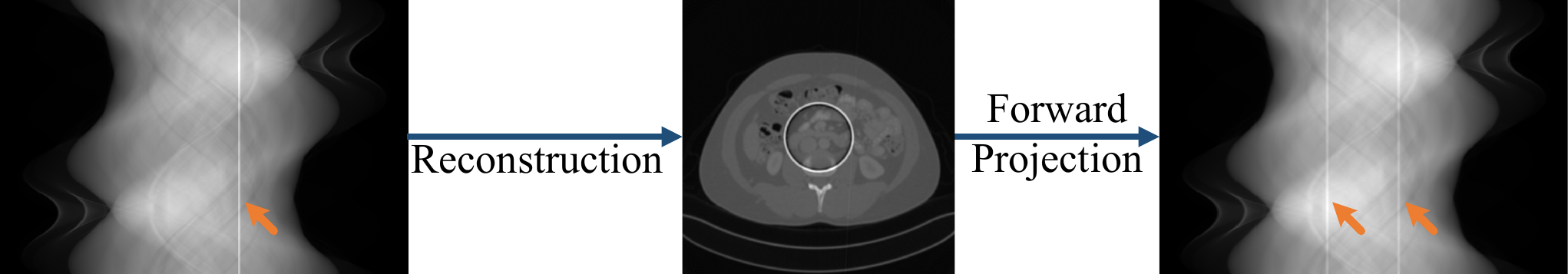}
  \caption{Illustration of how stipe artifacts in the projection image diffuse through reconstruction and forward projection processes.}\label{diff}
\end{figure}
Based on the theoretical analysis and key observations, we propose a ring artifacts correction method based on global-local features interaction guidance in the projection domain. We apply DWT to decompose the projection image into low-frequency sub-band that describe the global correlations of the projection and high-frequency sub-band that contain stripe artifacts. Considering the global correlation of projection images and the locality of stripe artifacts, we design the global and local features correction modules using VSS block and Dense block, respectively.
To more effectively localize and correct stripe artifacts, we innovatively introduce a global-local interaction guidance module, which incorporates an attention mechanism to guide and complement global and local features during the artifacts correction process. Additionally, we propose high-frequency fusion and reconstruction model in the encoding and decoding stages to achieve weighted fusion and reconstruction of multi-directional high-frequency sub-band. It should be noted that although image-domain methods have been widely studied, they lack the smoothness property under the Radon transform and the local response inconsistency characteristics in reconstruction images, making them unsuitable for the proposed modeling framework. While our method could be extended to a dual-domain framework, the high resource overhead and potential artifacts diffusion during alternating projection and reconstruction steps, shown in Fig.~\ref{diff}, motivate us to focus on the projection domain for ring artifacts correction.\par
The main contributions of this paper are as follows:
\begin{enumerate}
    \item We reveal the functional coupling between projection values and stripe artifacts through discrete wavelet decomposition and numerical analysis of real projection images, providing a new perspective for artifact modeling and correction.

    \item We employ the DWT to decouple the global correlations and stripe artifacts from the projection images, and design independent correction modules for global and local features using the VSS block and Dense block, respectively.

    \item Based on the functional coupling characteristics, we propose a global-local features interaction guidance module to facilitate the joint modeling and correction of global and local features.
\end{enumerate}
\hspace*{2em}The experimental results demonstrate that the proposed method achieves superior performance in CT image ring artifacts correction, with excellent visual quality and detail preservation. It provides both theoretical novelty and practical applicability, offering an effective and reliable solution for artifacts correction.

\section{Method}
\subsection{Framework Overview}
\begin{figure*}[ht]
  \centering
  \includegraphics[width=\textwidth]{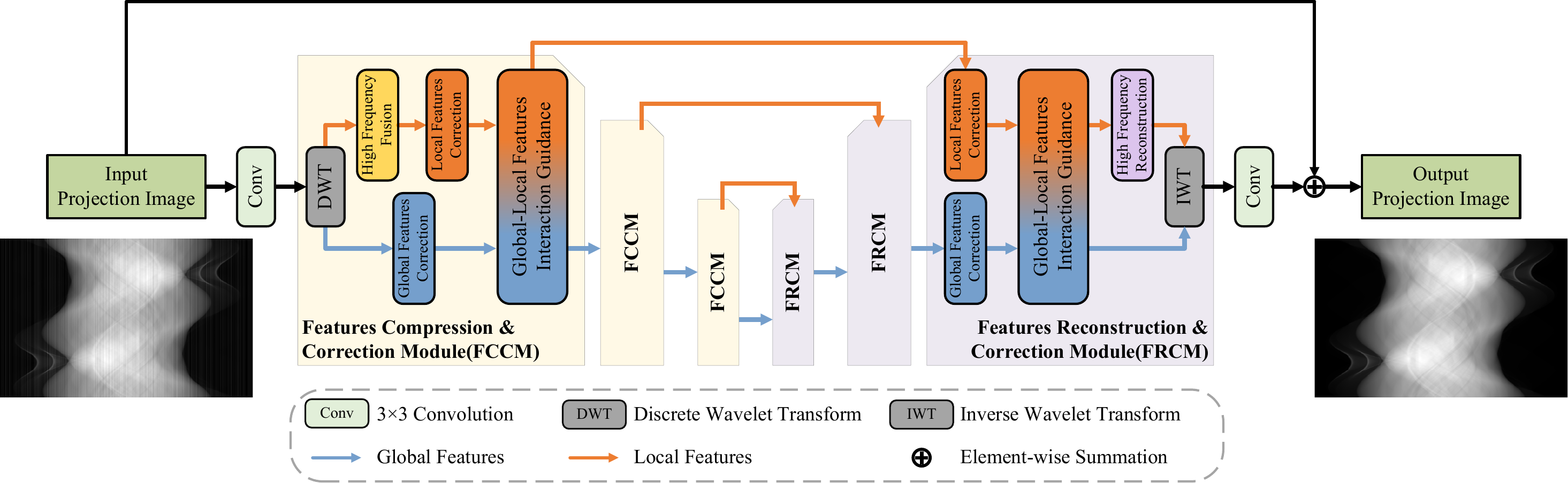}
  \caption{The overall structure of the proposed method. }\label{fig3}
\end{figure*}

To correct ring artifacts from CT reconstruction images effectively, we propose a network framework that models the global and local features of projection images in the projection domain. The proposed framework not only independently corrects the global and local features but also facilitates their collaborative optimization through a Global-Local Interaction Guidance module that integrates both, thereby enhancing the accuracy of artifacts correction. The structure of the network is shown in Fig.~\ref{fig3}. Based on the preceding analysis, we use Discrete Wavelet Transform (DWT) to perform downsampling, which allows effective separation of stripe artifacts from the global projection information. Inverse Wavelet Transform (IWT) is then used for upsampling, ensuring lossless reconstruction of the projection images. As Fig.~\ref{fig3} is shown, the proposed network comprises two key module: the Features Compression \& Correction Module (FCCM) and the Features Reconstruction \& Correction Module (FRCM), which correspond to the encoding and decoding phases of the network, respectively.\par
The FCCM effectively corrects both low-frequency and high-frequency features after DWT. For the local features branch, we first employ a High Frequency Fusion (HFF) module to perform weighted fusion of detail coefficients along the horizontal, vertical, and diagonal directions. The fused high-frequency features are then refined through a dedicated Local Features Correction (LFC) module. For the global features branch, a Global Features Correction (GFC) module is applied to restore the smoothness of the projection images. After the initial corrections in both branches, the global and local features are jointly fed into the Global-Local Features Interaction Guidance (GLFIG) module, which facilitates effective interaction between global and local representations. This design enables more accurate identification and suppression of localized stripe artifacts in the projection domain.\par
Based on the correction and interaction of global and local features, the FRCM introduces a High Frequency Reconstruction (HFR) module to reconstruct the corrected local features into horizontal, vertical, and diagonal sub-band required for IWT, which are then combined with the corrected low-frequency sub-band for IWT to accurately restore the projection image.

\subsection{Global and Local features correction modules}
The projection image is decomposed into low-frequency and high-frequency sub-band via DWT. The low-frequency sub-band mainly represent the background and smooth regions, reflecting the global correlations of the projection image, whereas the high-frequency sub-band include details such as edges and textures, which reflect the local nature of the stripe artifacts. Considering the distinct information carried by global and local features, we have developed independent correction modules for each.\par

\textbf{Global Features Correction (GFC)}:
The global features correction module employs the Visual State Space Block (VSS Block)~\cite{ruan2024vm} for modeling and correcting the global features. This module is constructed based on Mamba~\cite{liu2024vision}, which offers robust global modeling ability with linear computational complexity, enabling improved global information representation while efficiently controlling computational resource consumption.\par

\textbf{Local Features Correction (LFC)}:
In the local features correction module, we employ a local correction network consisting of multiple $3 \times 3$ small convolution kernels and dense residual connections~\cite{huang2017densely}, which aims to improve the model’s capability to model local details and stripe artifacts in the projection image. The small convolution kernels, with their smaller receptive fields, allow for accurate focusing on minute structural regions, while the dense residual connections help address the vanishing gradient problem and enhance features propagation efficiency, thereby boosting the model's ability to identify and correct stripe artifacts within the high-frequency sub-band.\par

\begin{figure*}[ht]
  \centering
  \includegraphics[width=\textwidth]{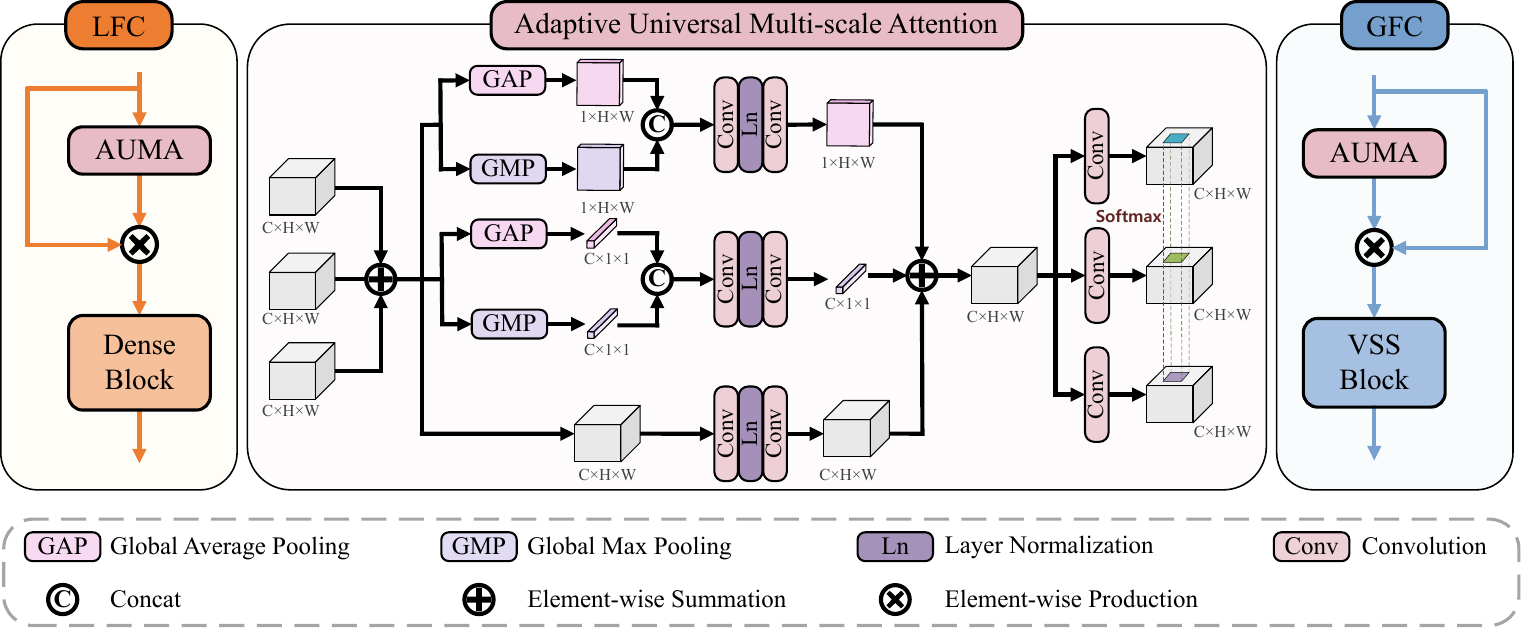}
  \caption{The structure of Global features Correction (GFC) module and Local features Correction (LFC) module. The middle shows the structure of Adaptive Universal Multi-scale Attention (AUMA) module.}\label{fig5}
\end{figure*}

\textbf{Adaptive Universal Multi-scale Attention (AUMA):}
In this study, we propose an Adaptive Universal Multi-scale Attention (AUMA) module, as illustrated in Fig.~\ref{fig5}. This module integrates Channel Attention, Spatial Attention, and Local Attention to enable dynamic feature weighting and adaptive modeling under various input compositions, thereby enhancing the module's capacity to capture multi-dimensional representations.
Given $N$ input feature maps $\boldsymbol{x}_1, \boldsymbol{x}_2, \dots, \boldsymbol{x}_N$, the AUMA module first performs element-wise summation to obtain the fused feature:
\begin{equation}
\boldsymbol{x} = \sum_{i=1}^{N} \boldsymbol{x}_i
\end{equation}
Subsequently, the fused feature $\boldsymbol{x}$ is processed through two internal branches within AUMA: a global features modeling branch $\mathcal{F}_{\text{Global}}(\cdot)$ and a local features modeling branch $\mathcal{F}_{\text{Local}}(\cdot)$, which collaboratively perform interaction-enhanced features modeling. Specifically, $\mathcal{F}_{\text{Global}}(\cdot)$ employs global average pooling (GAP) and global max pooling (GMP) along both the channel and spatial dimensions, while $\mathcal{F}_{\text{Local}}(\cdot)$ applies point-wise convolution to capture localized structural variations. The outputs of these two branches are summed and then passed through $N$ parallel convolution layers, followed by a Softmax normalization to produce $N$ attention weight $\boldsymbol{w}_i$ corresponding to each input:
\begin{multline}
\boldsymbol{w}_i = \text{Softmax}\left( \text{Conv}_i \left( \mathcal{F}_{\text{Global}}(\boldsymbol{x})
+ \mathcal{F}_{\text{Local}}(\boldsymbol{x}) \right) \right),\\  i=1,2,\dots,N
\end{multline}
The AUMA module is capable of dynamically allocating attention across multiple spatial scales. In the Global Features Correction (GFC) and Local Features Correction (LFC) modules, it enables adaptive weight generation based on the origin of each feature input, thereby improving the accuracy and robustness of ring artifacts modeling and correction. Moreover, AUMA supports multi-directional and multi-scale integration in tasks such as features interaction, high-frequency fusion, and reconstruction, demonstrating high flexibility and generalizability.

\subsection{Global-Local features Interaction Guidance Module}
\begin{figure}[ht]
  \centering
  \includegraphics[width=3.4in]{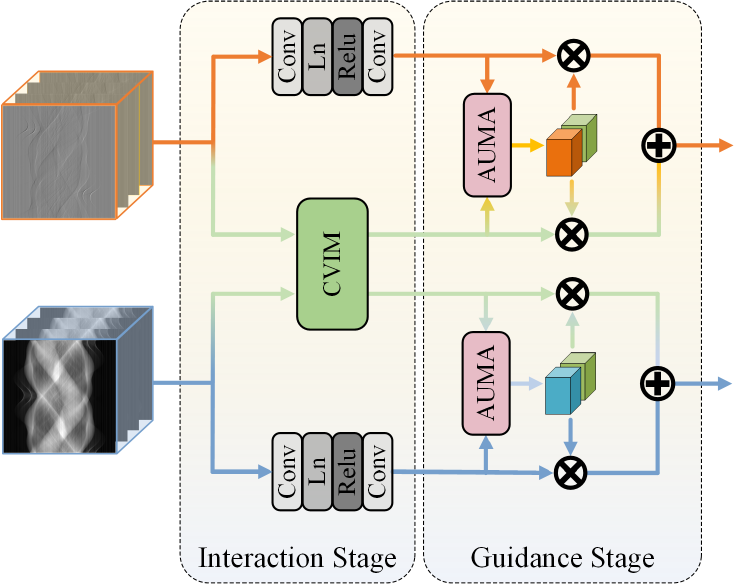}
  \caption{The structure of the Global-Local Features Interaction Guidance Module.}\label{fig6}
\end{figure}

After decomposing the projection image using DWT, the low-frequency sub-band capture the global correlation of projection values, while the high-frequency sub-band contain localized stripe artifacts. Due to the functional dependency between detector inconsistent responses and the projection values, these artifacts are typically coupled across both frequency sub-band. If the low-frequency and high-frequency sub-band are corrected independently and then directly reconstructed using IWT, the network’s ability to accurately model and suppress artifacts may be significantly constrained. To address this issue, we draw inspiration from the Cross-View Interaction Module (CVIM) proposed by Li \textit{et al.}~\cite{li2024multi} and design a \textbf{Global-Local features Interaction Guidance Module}, as shown in Fig.~\ref{fig6}. The module is composed of an interaction stage and a guidance stage, designed to facilitate information exchange between global and local features for their mutual refinement.\par
\textbf{Interaction stage}: The CVIM achieves complementary information fusion between left and right views using a cross-attention mechanism. Inspired by this, we adapt CVIM to enable information exchange between global and local features. When extracting local stripe artifacts, global features provide essential projection value context for accurately identifying amplitude variations. Likewise, during global features restoration, local features assist in locating residual artifacts regions, allowing for enhanced detail reconstruction and structure compensation. Serving as the core of the artifacts correction pipeline, this interaction stage depends on full cooperation between global and local features. The resulting interactive features, $\boldsymbol{I}_L^{\text{interact}}$ and $\boldsymbol{I}_H^{\text{interact}}$, represent a comprehensive fusion of both global and local information.\par
\textbf{Guidance stage}: Since the interactive features $\boldsymbol{I}_L^{\text{interact}}$ and $\boldsymbol{I}_H^{\text{interact}}$ represent only the outcome of global-local interaction and do not include the original features information, we further propose a guidance stage. In this stage, each interactive features is adaptively fused with its original counterpart to retain essential information while achieving mutual guidance. Specifically, we use AUMA to perform weighted fusion between $\boldsymbol{I}_L^{\text{interact}}$ and the original global features $\boldsymbol{I}_L$, and between $\boldsymbol{I}_H^{\text{interact}}$ and the original local features $\boldsymbol{I}_H$. The process is defined as follows:\par
\begin{align}
  \boldsymbol{I}_L^{\text{final}} & = \text{AUMA}(\boldsymbol{I}_L^{\text{interact}},\boldsymbol{I}_L) \\
  \boldsymbol{I}_H^{\text{final}} & = \text{AUMA}(\boldsymbol{I}_H^{\text{interact}},\boldsymbol{I}_H)
\end{align}
The proposed global-local features interaction guidance module facilitates cross-scale information interaction while preserving the representation of original structural features, thereby effectively enhancing the modeling capability for the spatial distribution and intensity variation of stripe artifacts.

\subsection{High-Frequency fusion and reconstruction module}
\textbf{High-Frequency Fusion (HFF)}:
In the encoding stage, the DWT decomposes the projection image into one low-frequency sub-band and three high-frequency sub-bands (i.e., $I_{LH}$, $I_{HL}$, and $I_{HH}$) corresponding to horizontal, vertical, and diagonal directions. These directional sub-band collectively represent the local details of the projection image and exhibit strong inter-correlation. To effectively integrate this information, we design a high-frequency fusion module that adaptively combines the three directional sub-bands using AUMA module. The fused high-frequency representation, denoted as $I_H$, not only reduces features dimensionality and computational overhead but also provides scale-aligned input for subsequent global-local guidance. Moreover, the AUMA module reinforces critical structural details during fusion, ensuring that key high-frequency features are retained despite compression.\par
\begin{figure}[ht]
  \centering
  \includegraphics[width=3.4in]{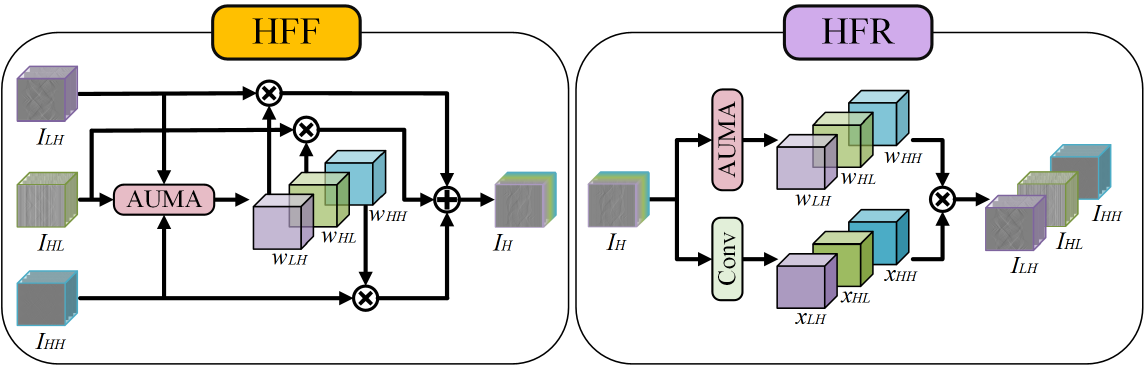}
  \caption{The structure of the High Frequency Fusion (HFF) Module and High Frequency Reconstruction (HFR) Module.}\label{fig7}
\end{figure}
\textbf{High-Frequency Reconstruction (HFR)}:
In the decoding stage, the IWT requires three sets of directional coefficients. To meet this requirement, we introduce a high-frequency reconstruction module that reverses the previous fusion process. Specifically, the fused high-frequency features $I_H$ is first decomposed into candidate features $x_{LH}$, $x_{HL}$, and $x_{HH}$, each corresponding to a wavelet sub-band. In parallel, direction-specific weight maps are generated and applied via Hadamard product to the respective features branches, resulting in the reconstructed sub-bands $I_{LH}$, $I_{HL}$, and $I_{HH}$. This design ensures precise restoration of the three directional sub-bands and provides high-quality structural details for the final projection image reconstruction.

\section{Experiment}
\subsection{Datasets Preparation}
We used the "2016 NIH-AAPM-Mayo Clinic Low-Dose CT Grand Challenge'' public dataset~\cite{chen2016mayo} for model training and testing, which contains a total of 5,936 CT slices acquired from 10 patients. Fan-beam circular scanning was simulated using the TIGRE toolbox~\cite{biguri2016tigre}, and Poisson noise with an average photon count of $1 \times 10^6$ was added to the simulated projections to generate projection data without stripe artifacts.\par
To simulate detector response inconsistency, we adopted the dataset preparation strategy proposed by Fang \textit{et al.}~\cite{fang2020removing}. Specifically, gain and offset factors were estimated from real detector measurements and applied to the stripe-free projection to generate simulated projections with ring artifacts. This perturbation process is formulated as:
\begin{equation}
p^{\text{simu}} = g' \cdot p^{\text{ref}} + o'
\end{equation}
where $p^{\text{ref}}$ denotes the projection without stripe artifacts, and $g'$ and $o'$ represent the sampled gain and offset factors from the statistical model. Subsequently, artifact-free and ring artifacts images were reconstructed using FBP with a Hamming window. All images were further transformed into polar coordinates to construct a multi-domain dataset across the image, projection, and polar domains.\par
The geometric parameters are set as follows: the source-to-detector distance (SDD) and the source-to-object distance (SOD) are 425~mm and 300~mm, respectively. The detector consists of 2048 elements with a spacing of 0.075~mm, and 384 projection angles were acquired per rotation.\par

\subsection{Implementation Details}
The proposed model was implemented using PyTorch and trained on a workstation equipped with an NVIDIA RTX 3090 Ti GPU (24\,GB memory). The input projection images had a resolution of $384 \times 2048$, and the reconstruction images were of size $512 \times 512$. The network consisted of four layers with a base channel number of $C = 64$. During training, we used the AdamW optimizer with parameters $(\beta_1, \beta_2) = (0.9, 0.999)$ and a weight decay of 0.02. The initial learning rate was set to $1 \times 10^{-3}$ and gradually decayed to $1 \times 10^{-5}$ using an exponential scheduler with decay factor $\gamma = 0.95$. The total loss function comprised two components: a pixel-wise loss and a perceptual loss between the network output and the reference image. The pixel-wise loss was computed using the L1 loss, while the perceptual loss was calculated as the mean squared error (MSE) between features extracted by a VGG19 network pretrained on ImageNet~\cite{johnson2016perceptual}. The weights assigned to the pixel-wise and perceptual losses were $\alpha_1 = 1$ and $\alpha_2 = 0.1$, respectively. The network was trained for 50 epochs.

\subsection{Competing Methods}
To ensure a fair and objective comparison, we compared the proposed method with three representative categories of ring artifacts correction methods.
\begin{enumerate}
    \item Image domain method: \textbf{Polar-TV}~\cite{yan2016variation} transforms the reconstruction images into the polar coordinate system and performs artifacts correction via an optimization-based regularization model, followed by inverse transformation back to the Cartesian space. In our experiments, we set the regularization parameters to $\lambda_1 = 0.0001$ and $\lambda_2 = 30$, with a maximum of 20 iterations. \textbf{ImgU-Net}~\cite{hein2023ring} is a U-Net-based correction framework operating in the image domain. It leverages perceptual loss to enhance ring artifacts suppression. All hyperparameters follow the original paper.
    \item Projection domain method: \textbf{WFF}~\cite{munch2009stripe} combines Fourier transform, wavelet decomposition, and Gaussian damping filters to suppress stripe artifacts in the projection domain. In our experiments, we use Daubechies 25 (DB25) wavelets with a decomposition level of 4 and a damping factor of 3. \textbf{MP-TVG}~\cite{wu2019removing} reduces ring artifacts by computing the mean of projection data and applying total variation (TV-L1) regularization along with Gaussian smoothing. The relaxation factor is set to 0.1 and the Gaussian standard deviation $\sigma = 0.5$. \textbf{ProjU-Net}~\cite{nauwynck2020ring} employs a five-stage U-Net architecture and integrates the unidirectional sparsity of projection data into the loss function to suppress stripe artifacts. Hyperparameters are configured according to the original settings.
    \item Dual-domain method: \textbf{Dual}~\cite{salehjahromi2019new} introduces sparse regularization in the projection domain and ring total variation (RTV) constraints in the image domain, solving an iterative optimization model for ring artifacts correction. We set $\lambda_1 = 0.1$, $\lambda_2 = 0.05$, and the maximum number of iterations to 30.
\end{enumerate}
To quantitatively evaluate the performance, we adopt three objective image quality metrics: Peak Signal-to-Noise Ratio (PSNR), Mean Squared Error (MSE), and Structural Similarity Index Measure (SSIM). For projection-domain and dual-domain methods, we also compare the quality of the corrected projection images. Higher PSNR and SSIM, as well as lower RMSE, indicate better performance.

\subsection{Experiments on Mayo 2016 Dataset}
\begin{figure*}[ht]
  \centering
  \includegraphics[width=\linewidth]{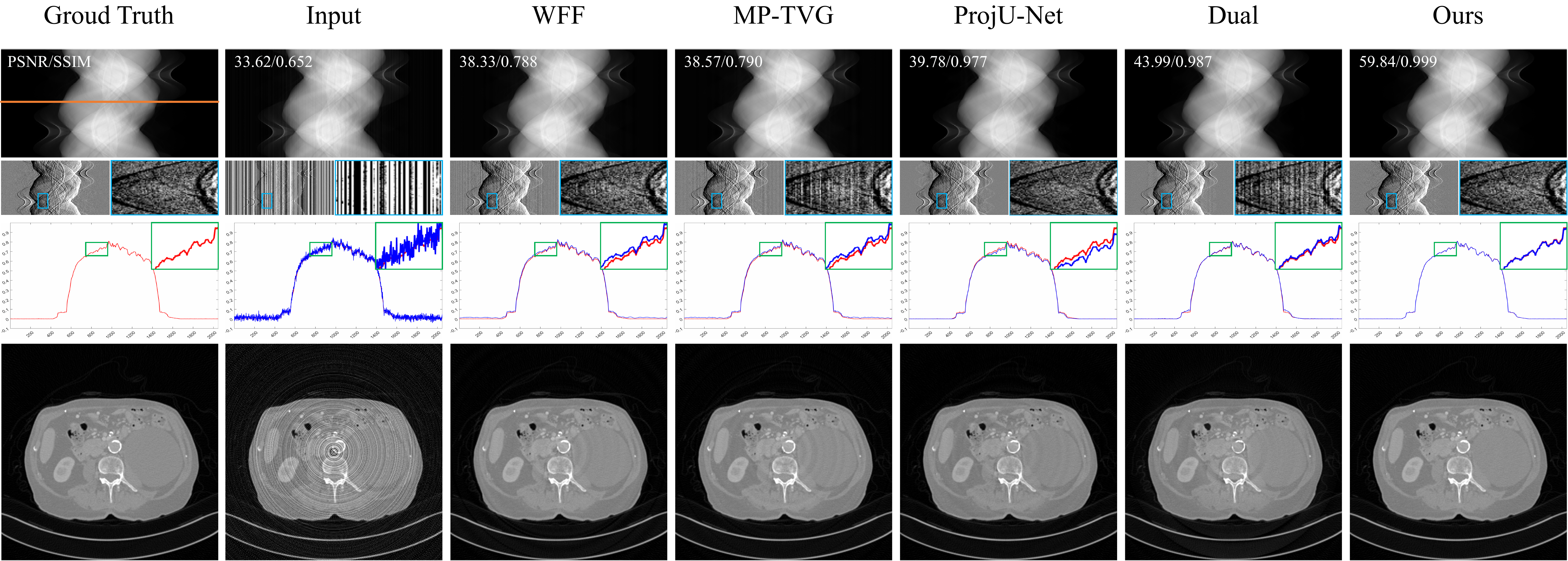}
  \caption{Qualitative results of projection domain methods and dual domain method from Mayo 2016 dataset.}\label{exp1}
\end{figure*}

In this section, we evaluate the performance of different methods in removing ring artifacts from CT images using the Mayo 2016 dataset. We first compare our method with projection domain and dual-domain correction methods. As shown in Fig.~\ref{exp1}, we present qualitative comparisons in terms of projection images, the vertical high-frequency sub-band of the projection’s wavelet transform and their detail images, central row gray-level profiles, and the Corresponding reconstruction images. The display ranges are set as follows: projection images \([0, 1.1]\), vertical high-frequency sub-band and detail images \([-0.005, 0.005]\), and reconstruction images \([0, 0.035]\).\par
As shown in the Fig.~\ref{exp1}, the WFF method suppresses part of the stripe artifacts, but residual artifacts with blurred patterns remain in both the vertical high-frequency subband and the reconstruction image. The MP-TVG method further weakens stripe artifacts, yet high-frequency residues and ring artifacts are still visible. The ProjU-Net method achieves better visual quality and structural recovery; however, slight deviations between projection values and labels are observed, with residual artifacts in the reconstruction image. The Dual method provides high projection fitting accuracy and detail restoration but fails to suppress high-frequency stripes effectively. In contrast, the proposed method significantly reduces low-frequency interference and effectively removes high-frequency stripe artifacts, producing reconstruction images with richer details and better consistency with the labels.\par
Quantitative results on projection images from Mayo 2016 dataset are summarized in Table~\ref{tab1}, where ours method achieves superior performance in PSNR, SSIM, and RMSE, demonstrating stronger artifact suppression and higher image fidelity.\par
\begin{figure*}[ht]
  \centering
  \includegraphics[width=\textwidth]{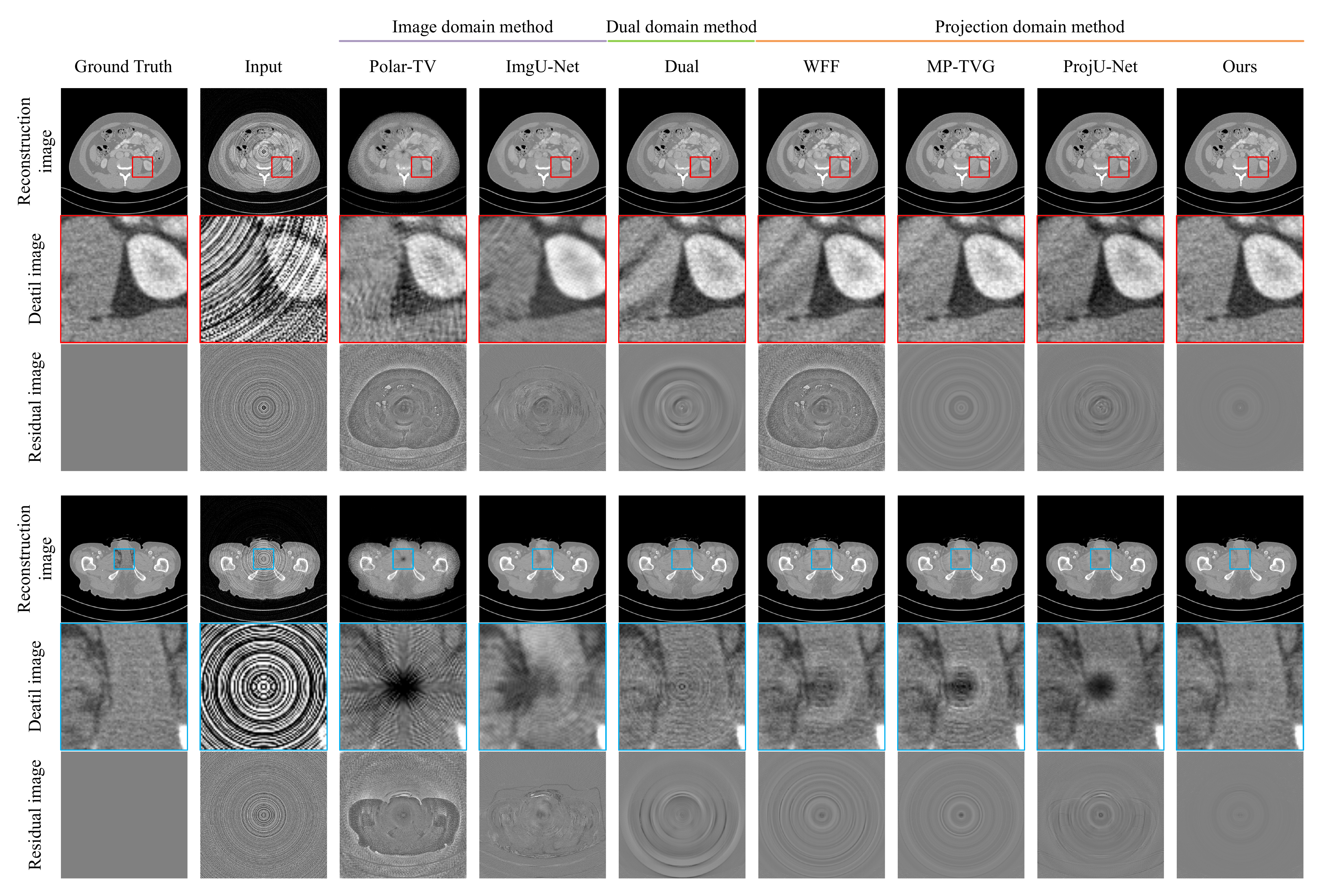}
  \caption{Qualitative results of all competing methods on reconstruction images from mayo2016 dataset.}\label{exp2}
\end{figure*}

\begin{table}[ht]
\centering
\caption{Quantitative result (MEAN$\pm$SD) of projection domain methods and dual domain method on projection images from Mayo 2016 dataset.}
\label{tab1}
\begin{tabular}{ccccc}
  \hline
  Method    & PSNR (dB)        & SSIM                & RMSE                \\ \hline
  Input     & 34.59$\pm$0.77  & 0.6867$\pm$0.0276   & 0.0220$\pm$0.0003    \\
  WFF       & 38.38$\pm$0.31  & 0.7906$\pm$0.0105   & 0.0134$\pm$0.0004    \\
  MP-TVG       & 38.58$\pm$0.32  & 0.7912$\pm$0.0105   & 0.0131$\pm$0.0004    \\
  ProjU-Net  & 40.68$\pm$1.01  & 0.9788$\pm$0.0035   & 0.0110$\pm$0.0011   \\
  Dual      & 43.36$\pm$0.46  & 0.9853$\pm$0.0014  & 0.0075$\pm$0.0004   \\
  Ours      & \textbf{58.85$\pm$2.11}     & \textbf{0.9997$\pm$0.0001}    & \textbf{0.0014$\pm$0.0003} \\\hline
\end{tabular}
\end{table}
We then further compared all competing methods on the reconstruction images. Fig.~\ref{exp2} shows two representative CT reconstruction slices processed by three categories of methods. The detail regions of these slices are marked by the red and blue boxes, respectively. The display ranges are configured as follows: reconstruction images \([0.005, 0.03]\), detail images \([0.015, 0.025]\), and residual images \([-0.01, 0.01]\).\par
As shown in Fig.~\ref{exp2}, the Polar-TV method enhances artifact spatial separability, but geometric errors from polar coordinate transformation and inverse transformation cause black spot artifacts in the image center, severely affecting image quality. ImgU-Net method shows some capability in suppressing strong artifacts but performs poorly in restoring edge details, with large errors. The Dual method offers a balance between structural and numerical fitting but fails to completely correct artifacts. In contrast, our method effectively eliminates ring artifacts while preserving fine details and accurately restoring low-frequency information. The residual images confirms that our method has the smallest error and highest accuracy. TABLE~\ref{tab2} provides the quantitative results of all methods on the reconstruction images from Mayo 2016 dataset, where our method consistently outperforms the second-best method in terms of PSNR, SSIM, and RMSE.\par
\begin{table}[ht]
\centering
\caption{Quantitative result (MEAN$\pm$SD) of all competing methods on reconstruction images from mayo2016 dataset.}
\label{tab2}
\begin{tabular}{ccccc}
  \hline
  Method    & PSNR (dB)          & SSIM               & RMSE (HU)   \\ \hline
  Input    & 21.85$\pm$2.14     & 0.3024$\pm$0.0448  & 200.350$\pm$6.719  \\
  Polar-TV & 22.99$\pm$1.71     & 0.5640$\pm$0.0215  & 98.764$\pm$2.081  \\
  ImgU-Net  & 35.18$\pm$1.82     & 0.8889$\pm$0.0161  & 40.450$\pm$3.780  \\
  Dual     & 35.38$\pm$0.62     & 0.9376$\pm$0.0043  & 33.919$\pm$1.408  \\
  WFF      & 42.43$\pm$1.06     & 0.9767$\pm$0.0035  & 17.390$\pm$1.390  \\
  MP-TVG      & 42.78$\pm$0.81     & 0.9742$\pm$0.0037  & 16.578$\pm$0.976  \\
  ProjU-Net & 41.16$\pm$2.16     & 0.9709$\pm$0.0043  & 23.997$\pm$1.526   \\
  Ours     & \textbf{50.84$\pm$2.41}     & \textbf{0.9977$\pm$0.0004}    & \textbf{7.193$\pm$1.306}  \\\hline
\end{tabular}
\end{table}

\subsection{Experiments on KiTS19 Dataset}
\begin{figure*}[ht]
  \centering
  \includegraphics[width=\linewidth]{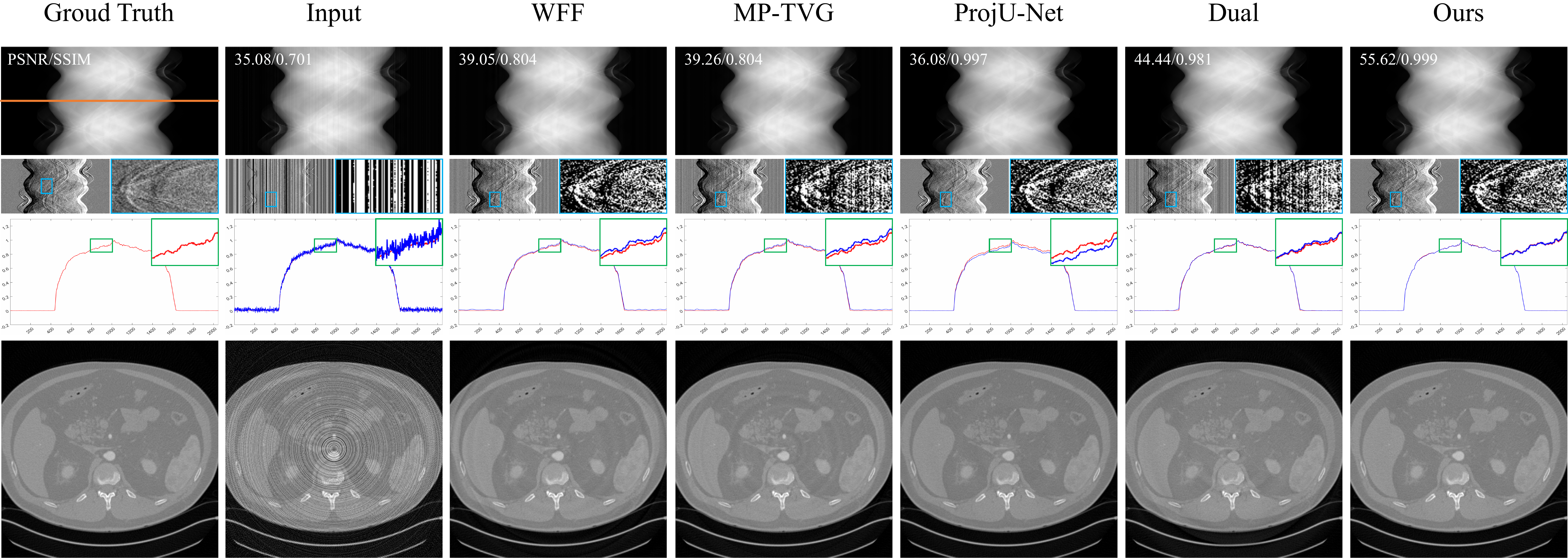}
  \caption{Qualitative results of projection domain methods and dual domain method from KiTS19 dataset.}\label{exp3}
\end{figure*}

In this section, we further validate the generalization performance of our method on the KiTS19 dataset to assess the robustness of different methods under dataset variations. Similar to the previous section, we first compare our method with projection domain and dual domain correction methods on the projection level. As shown in Fig.~\ref{exp3}, we present representative projection images, the vertical high-frequency sub-band of the projection’s wavelet transform and their detail images, central row gray-level profiles, and corresponding reconstruction images. The display ranges are set as follows: projection images \([0, 1.3]\), vertical high-frequency sub-band and detail images \([-0.005, 0.005]\) and \([-0.001, 0.001]\), and reconstruction images \([0, 0.035]\).\par
The results on the KiTS19 dataset are generally consistent with those on the Mayo 2016 dataset. Traditional projection domain methods (WFF and MP-TVG) show limited ability in suppressing high-frequency stripe artifacts, while ProjU-Net and Dual methods achieve better detail preservation and structural fidelity but still suffer from residual artifacts. In contrast, our method achieves the best performance in both artifact suppression and detail preservation. The profile plots further verify its superior fitting accuracy. As summarized in TABLE~\ref{tab3}, our method consistently outperforms other methods in terms of PSNR, SSIM, and RMSE.\par
\begin{figure*}[ht]
  \centering
  \includegraphics[width=\linewidth]{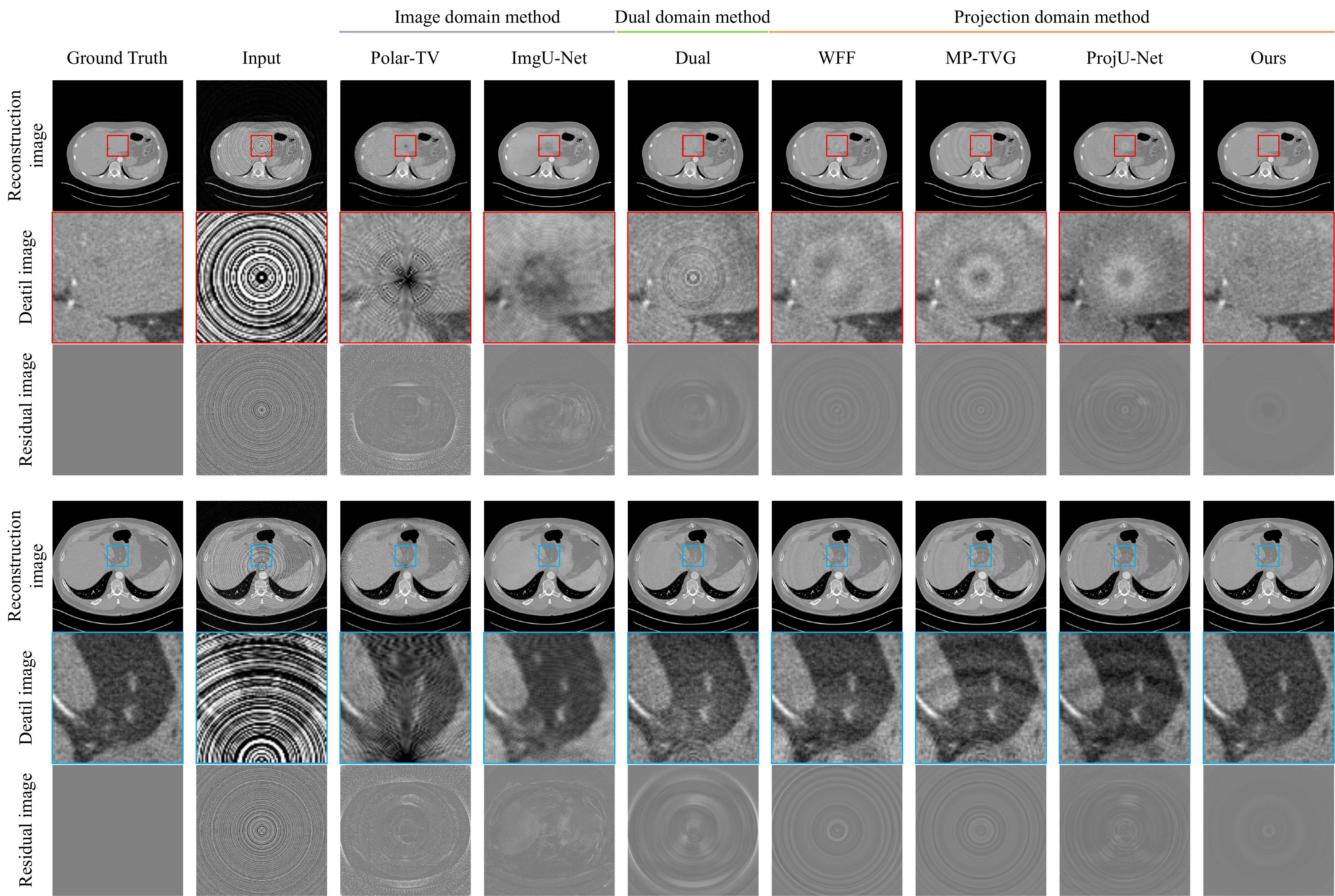}
  \caption{Qualitative results of all competing methods on reconstruction images from KiTS19 dataset.}\label{exp4}
\end{figure*}

\begin{table}[ht]
\centering
\caption{Quantitative result (MEAN$\pm$SD) of projection domain methods and dual domain method on projection images from KiTS19 dataset.}
\label{tab3}
\begin{tabular}{ccccc}
  \hline
  Method    & PSNR (dB)        & SSIM                & RMSE                \\ \hline
  Input     & 34.10$\pm$0.76  & 0.6518$\pm$0.0440   & 0.0220$\pm$0.0003    \\
  WFF       & 37.32$\pm$0.23  & 0.7335$\pm$0.0080   & 0.0134$\pm$0.0004    \\
  MP-TVG       & 37.50$\pm$0.24  & 0.7342$\pm$0.0081   & 0.0131$\pm$0.0003    \\
  ProjU-Net  & 39.96$\pm$1.32  & 0.9742$\pm$0.0141   & 0.0115$\pm$0.0027   \\
  Dual      & 43.06$\pm$0.36  & 0.9740$\pm$0.0025   & 0.0069$\pm$0.0003   \\
  Ours      & \textbf{58.28$\pm$1.49}     & \textbf{0.9993$\pm$0.0005}    & \textbf{0.0014$\pm$0.0002}  \\\hline
\end{tabular}
\end{table}
Fig~\ref{exp4} further illustrates the differences in reconstruction image quality across the three categories of methods. As in the previous section, we show two representative CT slices and their residual images for all competing methods and ours. The detailed regions of these slices are marked by the red and blue boxes, respectively. The display ranges are set as follows: reconstruction images \([0.005, 0.03]\), detail images \([0.015, 0.025]\), and residual images \([-0.01, 0.01]\).\par
The performance of different methods on the KiTS19 dataset is generally consistent with the results on the Mayo 2016 dataset. The Polar-TV method still suffers from black spot artifacts near the image center and shows limited effectiveness in artifact correction. The ImgU-Net and Dual methods achieve better structural restoration and detail preservation but still leave residual artifacts. In contrast, our method achieves better performance in both artifact suppression and detail recovery, especially in restoring fine details in the image center. As summarized in TABLE~\ref{tab4}, our method consistently outperforms other methods in terms of PSNR, SSIM, and RMSE.\par
\begin{table}[ht]
\centering
\caption{Quantitative result (MEAN$\pm$SD) on reconstruction image from KiTS19 dataset.}
\label{tab4}
\begin{tabular}{ccccc}
  \hline
  Method    & PSNR (dB)          & SSIM               & RMSE (HU)   \\ \hline
  Input    & 19.15$\pm$2.47     & 0.2517$\pm$0.0461  & 202.966$\pm$12.854  \\
  Polar-TV & 26.91$\pm$1.27     & 0.5597$\pm$0.0237  & 82.200$\pm$1.495  \\
  ImgU-Net  & 33.36$\pm$1.27     & 0.8449$\pm$0.0140  & 45.436$\pm$3.105  \\
  Dual     & 36.01$\pm$0.95     & 0.9378$\pm$0.0044  & 30.159$\pm$1.162  \\
  WFF      & 37.87$\pm$0.74     & 0.9001$\pm$0.0058  & 26.650$\pm$0.682  \\
  MP-TVG      & 37.67$\pm$0.65     & 0.8923$\pm$0.0051  & 27.112$\pm$0.593  \\
  ProjU-Net & 40.10$\pm$1.50     & 0.9630$\pm$0.0092  & 22.689$\pm$2.513   \\
  Ours     & \textbf{49.96$\pm$1.57}     & \textbf{0.9966$\pm$0.0008}    & \textbf{6.568$\pm$1.047}  \\\hline
\end{tabular}
\end{table}

\subsection{Experiments on Real data}
To further evaluate the effectiveness and robustness of the proposed method in practical scenarios, we conducted experiments using two real biological samples: pig hoof and pork. The experiments were performed using a laboratory-built imaging platform, equipped with a HAMAMATSU-L10101 X-ray source and an EIGER2 1MW R-DECTRIS PCD. The scanning geometry was consistent with that used in the simulation experiments. For simplicity, we selected the central slice from each dataset and simulated the fan-beam CT imaging process using its corresponding projection data.\par
\begin{figure*}[ht]
  \centering
  \includegraphics[width=\linewidth]{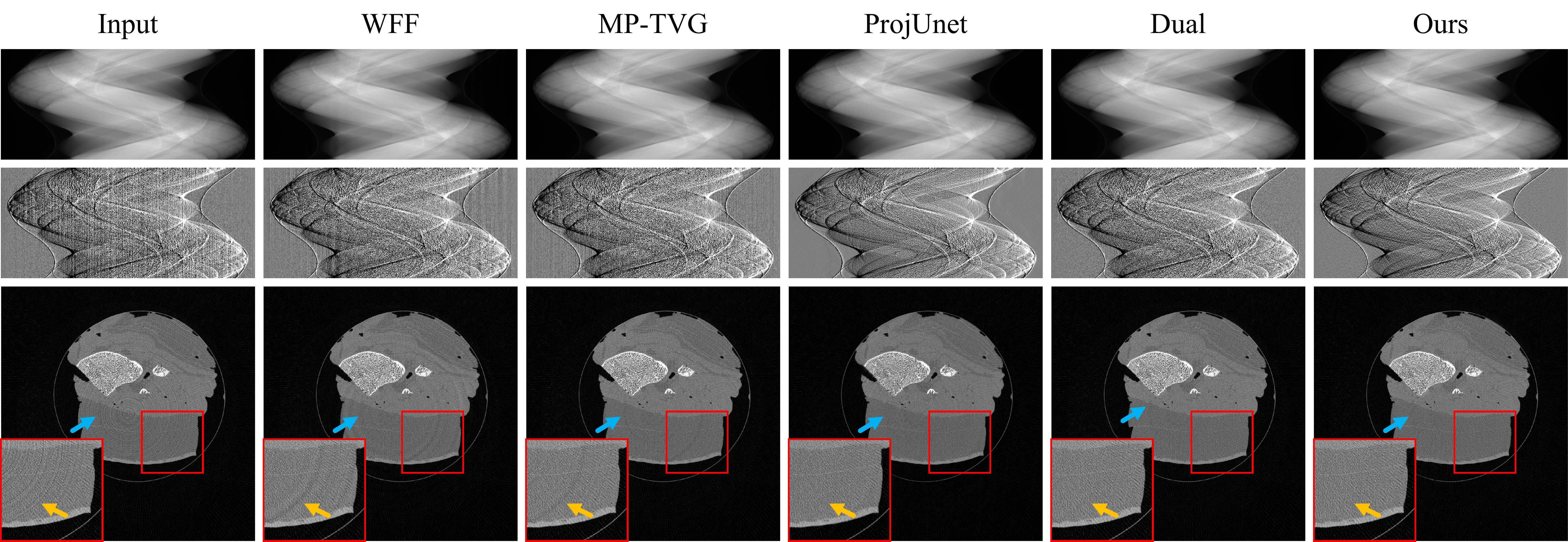}
  \caption{Qualitative results of projection domain methods and dual domain method from real experiment.}\label{exp5}
\end{figure*}

We first test the real data from the pork sample and compare our method with typical projection domain methods (WFF, MP-TVG, ProjU-Net) and dual domain methods (Dual). The results are shown in Fig.~\ref{exp5}.
The Fig.~\ref{exp5} shows the processed projection images, the corresponding vertical high-frequency sub-band and their detail views, and the reconstruction images. The detail regions are marked by the red box and the blue and orange arrows. The display ranges are: projection images [0,1.4], vertical high-frequency sub-band and detail images [-0.005,0.005], reconstruction images [0,0.06] and detail images [0,0.05].\par
The image reconstructed from the real projection data using FBP show that ring artifacts span across multiple tissue structures (fat, bone, etc.), interfering with image quality. Comparison results indicate that WFF method reduces artifacts but leaves significant stripe residues in high-frequency details. The artifacts are not fully suppressed in the reconstruction image. The MP-TVG method moderately suppresses ring artifacts but introduces diffused artifacts that impair local image clarity. The ProjU-Net method effectively restores projection structures and corrects most stripe artifacts, yet ring-like patterns remain in the reconstruction image. The Dual method shows good performance in structural restoration and artifact suppression, offering better visual quality overall, although slight residuals are still visible. In comparison, our method not only restores low-frequency global structure but also successfully corrects stripe artifacts from the vertical high-frequency details, preserving image details and improving the overall quality of the reconstruction image.\par
\begin{figure*}[ht]
  \centering
  \includegraphics[width=\linewidth]{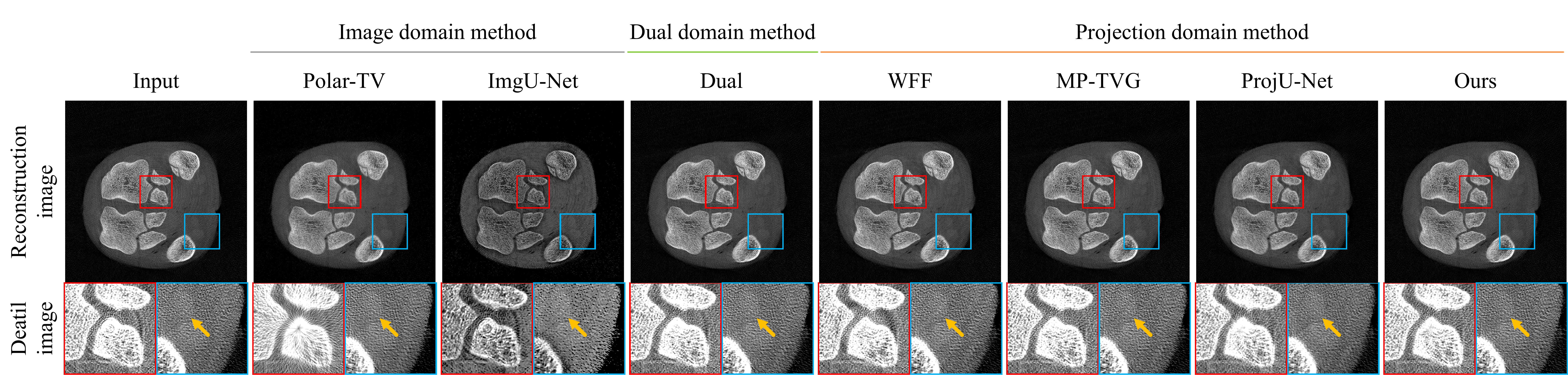}
  \caption{Qualitative results of all competing methods on reconstruction images from real data.}\label{exp6}
\end{figure*}

We then evaluated the performance of three categories of methods on the pig trotter sample, which presents more complex tissue structures. Fig.~\ref{exp6} shows the reconstruction images of all competing methods, along with the zoom-in views of two detail regions marked by the red and blue boxes. The display ranges were set to \([0, 0.07]\) for the reconstruction images and \([0, 0.04]\) for the zoom-in views of detail images.\par
As shown, although traditional methods suppress ring artifacts overall, they struggle to capture complex tissue structures and boundaries in low-SNR real data. This leads to noticeable artifact residues and diffusion in detailed regions, such as the central area, bone edges, and tissue interfaces. Local edge blurring is also evident. Image domain methods fail to effectively correct ring artifacts, with issues like numerical deviations and inaccurate structural recovery. Projection domain deep learning methods perform slightly better in detail retention but still lack complete artifact elimination and central structure recovery. In contrast, our proposed method achieves superior structural fidelity and robust artifact suppression under complex tissue conditions, especially excelling in the recovery of central details. These results demonstrate its strong generalization capability and potential for real applications.

\subsection{Ablation study}
To further assess the effectiveness of the Global-Local features Interaction Guidance (GLFIG) module, we constructed an ablation model (denoted as w/o GLFIG), which independently corrects global and local features without GLFIG module. to guarantee methodological parity, the network architecture and parameter settings of the w/o GLFIG were kept consistent with ours model proposed in this paper. We trained and tested the w/o GLFIG on the Mayo 2016 dataset, and the TABLE~\ref{tab_ablation} demonstrate that all the proposed module significantly contributes to artifact correction. However, when the GLFIG module was removed, average PSNR decreased from 58.85 dB to 55.48 dB, average SSIM dropped from 0.9997 to 0.9983, and RMSE increased from 0.0014 to 0.0020, indicating a decline in overall performance.\par
\begin{table}[ht]
\centering
\caption{Quantitative result (MEAN$\pm$SD) of ablation study on projection image from Mayo 2016 dataset.}
\label{tab_ablation}
\begin{tabular}{cccc}
  \hline
  Method   & PSNR (dB)          & SSIM               & RMSE (HU)   \\ \hline
  w/o GLFIG  & 55.48$\pm$1.53  & 0.9983$\pm$0.0011  & 0.0020$\pm$0.0003 \\
  Ours      & 58.85$\pm$2.11     & 0.9997$\pm$0.0001    & 0.0014$\pm$0.0003\\
  \hline
\end{tabular}
\end{table}
\begin{figure}[ht]
  \centering
  \includegraphics[width=3.4in]{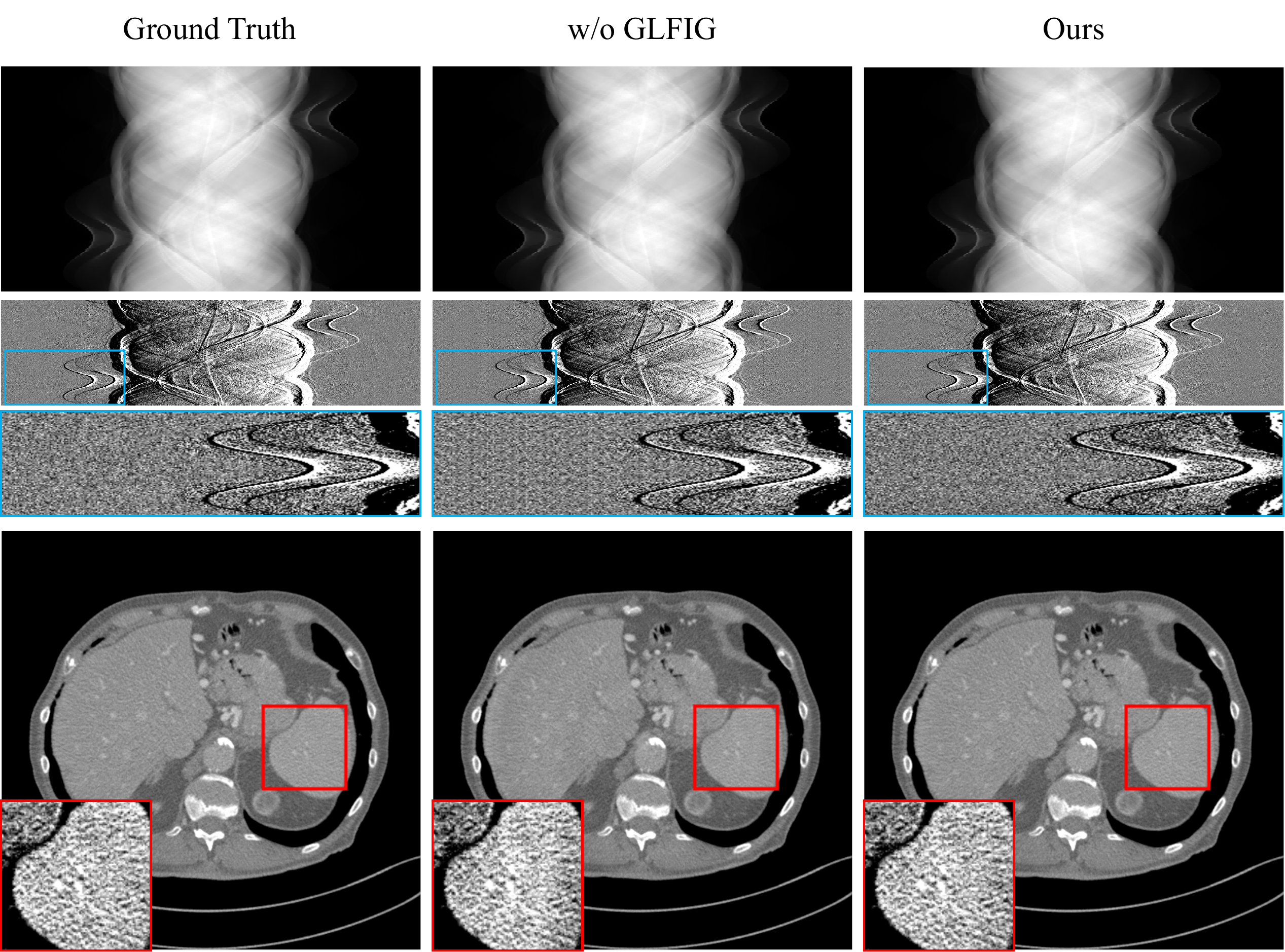}
  \caption{Qualitative results of ablation study from Mayo 2016 dataset.}\label{exp_ablation}
\end{figure}
We selected a representative projection image for visual analysis, as shown in Fig.~\ref{exp_ablation}. Each row presents the projection image, the vertical high-frequency detail coefficient image, its zoomed-in image, and the corresponding reconstruction image with detail images. The display ranges are [0, 1.1] for the projection images, [-0.005, 0.005] and [-0.0015, 0.0015] for the vertical high-frequency sub-band and detail images, and [0.01, 0.03] and [0.02, 0.023] for the reconstruction images and their detail images, respectively. While both correct stripe artifacts in the projection images, further analysis of the vertical high-frequency sub-band shows that the \textit{w/o GLFIG} model retains visible stripe artifacts, particularly in the background and along edges. The corresponding reconstruction images and their detail images further support this conclusion. In the smoother areas of the reconstruction image, the \textit{w/o GLFIG} model exhibits noticeable ring artifact residues. In contrast, the complete model significantly outperforms the simplified model in terms of artifact suppression, structural clarity, and detail restoration.\par
The above experiment results clearly validate the importance of Global-Local features Interaction Guidance module in the proposed network architecture. By facilitating effective interaction and joint optimization of global and local features, this module greatly improves artifact suppression and the quality of the reconstruction images.

\section{Discussion}
In this study, we proposed a ring artifacts correction method for CT images based on global-local features interaction guidance in the projection domain. We first analyzed the functionally coupled relationship between the projection images and stripe artifacts through wavelet transform analysis. Then, the projection images were decomposed into low-frequency sub-band reflecting global smoothness and high-frequency sub-bands containing local disturbances using discrete wavelet transform, and corresponding artifact correction modules were designed respectively according to the characteristics of global and local features. Ablation experiments demonstrated that the proposed independent correction modules, even without features interaction guidance, achieved superior ring artifacts correction performance compared to most existing methods, verifying the effectiveness and rationality of the global and local features correction modules. However, independent correction still exhibited certain performance limitations. With the introduction of the global-local features interaction guidance module, the information flow and collaborative optimization between global and local features were effectively enhanced, further improving the artifacts correction performance. Experimental results also confirmed that the global-local features interaction guidance module plays a critical role in boosting the overall performance of the proposed model.\par
Although the proposed method demonstrates excellent performance in artifacts correction and image restoration, there are still some limitations. First, the effectiveness of the method has not yet been evaluated under extreme conditions such as detector units failure, where the artifacts become more complex and severe information loss may occur. In future work, more prior knowledge from projection or reconstruction images will be introduced to enhance the robustness of the model under such challenging scenarios. Second, deep learning-based artifacts correction methods typically rely on sufficient and high-quality training data. To address this issue, we will explore extending the proposed method to semi-supervised or unsupervised learning frameworks, enabling the model to achieve comparable artifacts correction performance with limited labeled data.

\section{Conclusion}
This study reveals the functional coupling between stripe artifacts and dynamic variations of projection data through wavelet domain analysis, and innovatively introduces the concept of global-local feature interaction guidance for stripe artifacts correction. Specifically, we propose a CT ring artifacts correction method based on global-local features interaction guidance in the projection domain. Utilizing discrete wavelet transform, the proposed approach decomposes projection image into low-frequency components, which reflect global correlations, and high-frequency components containing local disturbances. A network framework is then designed to independently correct these global and local features while promoting their mutual interaction. The proposed method demonstrates outstanding performance in CT ring artifacts correction, as validated by experiments on two open-source datasets and real data, effectively suppressing artifacts and significantly improving image quality. Ablation experiments further confirm the efficacy of our method.

\section*{Acknowledgments}
The authors are grateful to the Beijing Higher Institution Engineering Research Center of Testing and Imaging as well as the Beijing Advanced Innovation Center for Imaging Technology for funding this research work.

\bibliography{references}
\bibliographystyle{IEEEtran}

\end{document}